\documentstyle[preprint,aps]{revtex}
\begin{document}
\include{epsf}
\draft

\title{\Large \bf Degenerate fermion gas heating by hole creation}
\author{
Eddy Timmermans\\
 {\small Theoretical Division (T-4), Los Alamos National Laboratory,} 
{\small Los Alamos, NM 87545}}
\maketitle
\begin{abstract}

	Loss processes that remove particles from an 
atom trap leave holes behind in the single particle
distribution if the trapped gas is a degenerate fermion system.  
The appearance of  
holes increases the temperature and we show that the heating is (i) 
significant if the initial temperature is well below the 
Fermi temperature $T_{F}$, and (ii) increases the temperature
to $T \geq T_{F}/4$ after half of the system's lifetime,
regardless of the initial temperature.  The hole heating
has important consequences for the prospect of observing 
Cooper-pairing in atom traps.

\end{abstract}

\pacs{PACS numbers(s):03.75.Fi, 05.30.Jp, 32.80Pj, 67.90.+z}


        Insofar as the ultra-cold atoms behave as 
an equilibrium system, the neutral atom trap provides 
a laboratory for the study of low-temperature phenomena. 
Recent experiments with atomic Bose-Einstein
condensates (BECs) \cite{becs} further highlight
this connection by reporting genuine superfluid
behavior \cite{super}. Similarly, the observation of 
fermion pairing in the gas phase would be of equal, or 
even greater interest \cite{leggett}, and the quest 
for atom-trap fermion superfluidity has begun in
earnest.  In this Letter, we 
discuss the effects of particle-loss to which
the metastable ultra-cold atom gases are prone. In 
addition to limiting the system's lifetime, $\tau_{L}$,
the loss processes leave behind holes in the single 
particle distribution of a degenerate fermion gas
(of temperature $T$ significantly lower
than the Fermi-temperature $T_{F}$).  The fermions heat
in the subsequent relaxation and, without additional cooling,
the system's temperature doubles from its initial
value $T_{0}$ within a time $\tau_{2} \sim 20 
[T_{0}/T_{F}]^{2} \tau_{L}$.
After $t \sim 0.5 \tau_{L}$ the fermion gas heats up
to $T \geq 0.25 T_{F}$, regardless of the initial temperature.
The competition of superfluid formation
with `Fermi-hole heating' leads to stringent lower-bounds 
for the strength of the inter-atomic attraction that 
would pair the fermions.

        \underline{Introduction}
In atom traps, inelastic collisions 
impart kinetic energy to the scattering products, which 
subsequently leave the trap. Scattering with
background atoms and inelastic two-and three-body collisions 
decrease the local particle density, $n$, as $\dot{n} 
= - \gamma n, \dot{n} = - \beta n^{2}, \dot{n} = - \alpha n^{3}$, 
respectively.  Although two and three-body scattering 
is suppressed in a single fermion gas,
so are the inter-atomic interactions that relax
the system to thermal equilibrium.  By trapping 
$^{40}K$-atoms in two different internal states 
D. Jin's group circumvented the problem and succeeded 
in cooling fermion atoms evaporatively
(i.e. by removing the highest energy atoms) to degeneracy \cite{Jin}. 
Most recently, R. Hulet's group at Rice University \cite{huletscience}
and C. Salomon's group at the ENS, Paris \cite{salomon},  
cooled fermion atoms ($^{6}$Li) by bringing them
in thermal contact with ultra-cold ($^{7}$Li)--bosons. 
The lowest reported temperatures \cite{huletscience} are $T \sim 0.25 T_{F}$,
and the fermion mixture cooling efforts appear to have encountered
a limit in this temperature range \cite{Jin}.  
Calculations suggest that Pauli-blocking 
plays a role \cite{mholland} but this effect alone does not seem to explain 
the limit.  An alternative motivation for creating
fermion gas mixtures is the prospect of s-wave Cooper-pairing.
This scheme -- the most likely mechanism to
achieve fermion superfluidity at realistic atom trap densities,
$n \sim 10^{12} - 10^{15} cm^{-3}$ -- requires the
s-wave interaction between two types of distinguishable atoms 
to be attractive
(i.e. described by a negative scattering length, $a < 0$).  Whether the
atom mixtures are created to equilibrate the system 
or to form Cooper-pairs, the
unlike fermion interactions introduce loss processes.

	We consider a weakly interacting gas mixture of
the same fermion isotope in two different internal states, 
$1$ and $2$.  The optimal mixture contains 
equal particle densities $n_{1} = n_{2} = n$, 
corresponding to equal particle numbers, $N_{1} = N_{2} = N$
\cite{hulet1}.  The usual trapped neutral atom gases are
dilute in the sense that $|a|/r << 1$ where $r$ denotes 
the average inter-particle distance,
$r = n^{-1/3}$ (typically, $|a| \sim 1 {\rm nm}$ whereas
$r \sim 1. - 0.1 \mu{\rm m}$).  In contrast, the quest for fermion
superfluidity necessitates an unusually strong
neutral atom attraction.  This condition follows from
the critical temperature $T_{c}$ for s-wave Cooper-pairing:
in the described homogeneous fermion mixture of
Fermi-momentum $k_{F} = (6\pi^{2})^{1/3} r^{-1} \approx 3.9 r^{-1}$, 
Fermi-energy $\epsilon_{F} = \hbar^{2} k_{F}^{2}/2m$, and Fermi-temperature
$T_{F} = \epsilon_{F}/k_{B}$ (where $k_{B}$ is the Boltzmann
constant) $T_{c}$ equals \cite{gorkov}
\begin{equation}
T_{c} \approx 0.3 T_{F} \exp^{-\frac{\pi}{2 k_{F} |a|}} =
0.3 T_{F} \exp^{-0.4 \frac{r}{|a|}} \; .
\label{e:hh1}
\end{equation}
With $T_{F} \approx 3.65 \mu{\rm K} A^{-1} (n/n_{0})^{2/3}$, where
$n_{0}$ is a realistic reference density, $n_{0} = 10^{12} cm^{-3}$, and
$A$ the mass number, we find that the prefactor of (\ref{e:hh1}) 
is accessible to present day atom trap technology, but
with $|a|/r \sim 10^{-3}$,
$T_{c} \approx 0.3 T_{F} \exp(-400)$.  

	\underline{Lifetime restrictions}	
Even if such exponentially
low temperatures can be reached, the time scale on which
the superfluid forms, $\tau_{form}$, exceeds any realistic 
lifetime $\tau_{L}$.
This formation time, $\tau_{form}$, depends on the
Fermi-time scale, $\tau_{Fermi} = \hbar/\epsilon_{F} 
\approx 2.09 \mu{\rm sec} A (n/n_{0})^{-2/3}$ and takes on the
form \cite{Houbiers}
\begin{equation}
\tau_{form} \sim \frac{\tau_{Fermi}}{2\pi} \left[ 
\frac{T_{F}}{T_{C}} \right]^{2}
\approx \frac{10}{2\pi} \tau_{Fermi} \exp(0.8  r/|a|) .  
\label{e:hh1b}
\end{equation}
The requirement that the superfluid forms before the gas is depleted, 
$\tau_{form} < \tau_{L}$, constrains the scattering length according to
$|a|/r > 0.8 / [ ln(2\pi/10) + ln(\tau_{L}/\tau_{Fermi})] \approx 0.35/
log_{10}(\tau_{L}/\tau_{Fermi})$ (assuming $\tau_{L} >> 10^{2} \tau_{Fermi}$).
For typical values, $\tau_{Fermi} \sim 10 - 100 \mu{\rm sec}$
and $\tau_{L} \sim 100$ seconds, $log_{10}(\tau_{L}/\tau_{Fermi})
\sim 6 - 7$, we find that the scattering length has to exceed
$5 \%$ of the average inter-particle distance, $|a|/r \geq 0.05 - 0.06$.  
Atom gas Cooper-pairing thus requires an effective
interaction that is unusually strong for neutral
atoms.  Such interaction can be realized by using atoms that form
a virtual state in the binary scattering,
or by using external fields that alter the atom-atom interactions.
On the other hand, when the magnitude of the
negative scattering length exceeds $ |a| > 0.48 r$ (corresponding
to $k_{F} |a| > 3 \pi/5$, \cite{hulet1}), the pressure turns negative
and the gas collapses.  This requirement of mechanical stability
limits the critical Cooper-pairing temperature to $T_{c} < T_{c,max} 
\approx 0.1 T_{F}$.

\underline{Hole heating}  Next, we illustrate 
the heating caused by the loss-induced
creation of fermi-holes.  We consider a homogeneous fermion
mixture  that is `normal' (i.e. not superfluid) and initially 
at zero temperature. The single particle distributions
are filled fermi-spheres in momentum space, corresponding
to occupation numbers $n_{j,{\bf k}} = 1$ ($j=1,2$) if $k \leq k_{F}$,
$n_{j,{\bf k}} = 0$ if $k > k_{F}$.  The loss processes `perforate'
the fermi-spheres, creating holes that bring the system into a
state that is not the ground-state (which has filled
fermi-spheres of reduced radius).  In due course, the system 
relaxes to its thermal equilibrium at finite temperature.
The scattering processes that relax 
the system are interesting in their own right. A hole in 
Fermi-sphere 1 is filled by a particle of initially higher energy.  This 
fermion 1 can change its momentum by interacting with a fermion 
2 that is thereby promoted to an energy above the Fermi-level.  Such 
particle-hole scattering,  akin to traditional Auger scattering,
continuously produces particles with energies up to $2 \epsilon_{F}$, 
energies that are anomalously high from the perspective of a 
thermal distribution. The resulting `high-energy' fermions are 
scattered to lower energy states (or `cooled').

        The importance of hole heating follows from the rate 
of temperature increase which depends on the specific heat per
particle, $c_{V} = (\pi^{2}/2) k_{B} [k_{B} T / \epsilon_{F}]$ 
and on the rate with which particles are removed,
\begin{equation}
\dot{n}_{j,{\bf k}} = - \frac{1}{\tau_{L}} n_{j,{\bf k}} \; ,
\label{e:hh2}
\end{equation}
where the system's lifetime, $\tau_{L}$, can depend on the
density ($\tau_{L}^{-1} = \gamma, \beta n, \alpha n^{2}$ for background,
two- and three-body loss respectively) \cite{remark}.  
The removal of a single particle
of momentum ${\bf k}$ lowers the energy of $N$ fermions at temperature 
$T$, $E(N,T)$, by its kinetic energy but increases the energy 
relative to that of the remaining particles, $N-1$ in total, 
at the same temperature,
\begin{eqnarray}
E' &=& E(N,T) - \frac{\hbar^{2} k^{2}}{2m} 
\nonumber \\
&\approx&
E(N-1,T) + \left[ \epsilon_{F} - \frac{\hbar^{2} k^{2}}{2m} \right] \; ,
\label{e:hh3}
\end{eqnarray}
where we assumed $T << T_{F}$ so that
the chemical potential is approximately equal to the Fermi-energy:
$E(N,T) - E(N-1,T) \approx \epsilon_{F}$.  Therefore, upon removal
of a particle of momentum ${\bf k}$, the system `gains' 
$[\epsilon_{F} - \hbar^{2} k^{2} /2m]$ in excess to the energy
of the remaining particles at the same temperature.  The probability
to create a hole in the $(j,{\bf k})$--state after a short time interval 
$\Delta t$ is equal to $- \Delta t \; \dot{n}_{j,{\bf k}} 
= \frac{\Delta t}{\tau_{L}}
n_{j,{\bf k}}$.
The total excess energy gathered by
the system during $\Delta t$ is the ($j,{\bf k}$)-
excess energy multiplied by the probability for hole creation,
summed over all ($j,{\bf k}$)--states:
\begin{eqnarray}
\Delta E &=& \left( \frac{\Delta t}{\tau_{L}} \right)
\sum_{j,{\bf k}} n_{j,{\bf k}} \left[ \epsilon_{F} -
\frac{\hbar^{2} k^{2}}{2m} \right] 
\nonumber \\
&=& \left( \frac{\Delta t}{\tau_{L}} \right) \;
\frac{4}{5} N \epsilon_{F} \; .
\label{e:hh5}
\end{eqnarray}
The corresponding temperature increase $\Delta T = \Delta E / [2 N c_{V}]$
gives a rate, $\dot{T} = \Delta T / \Delta t$, equal to
\begin{equation}
\dot{T} = \frac{\Delta E}{\Delta t C_{V}} = \left( \frac{4}{5\pi^{2}} 
\right)
\; \frac{T_{F}^{2}}{\tau_{L} T} \; .
\label{e:hh6}
\end{equation}
The solution to Eq.(\ref{e:hh6}), parametrized by the initial
temperature $T_{0}$ and the `temperature doubling time',
$\tau_{2}$,
\begin{equation}
T = T_{0} \sqrt{ 1 + \frac{ 3 t}{ \tau_{2} } } 
\; \; ;
\; \tau_{2} = \left( \frac{15 \pi^{2}}{8} \right) \; 
\left[ \frac{ T_{0} }{T_{F}} \right]^{2} \; \tau_{L} \; ,
\label{e:hh7}
\end{equation}
shows that the system doubles its temperature in a time
that is proportional to the square of the Fermi to initial
temperature ratio, $\tau_{2}
\sim 20 [T_{0}/T_{F}]^{2} \tau_{L}$.
Thus, fermions brought to $1\%$ of their Fermi-temperature
double their temperature in $\sim 0.2 \%$ of their lifetime.

        The above derivation assumes a constant loss-rate, whereas
a change in the density alters this rate on a long enough time
scale.  Also, the use of the specific heat, a statistical quantity,
may give the impression that the validity of the above 
derivation requires the system to remain in thermal equilibrium.
To address both issues we determine the temperature
from the observation that the
energy-per-particle, $e$, of a homogeneous fermion gas
or fermion gas mixture, is not affected by the particle loss
(\ref{e:hh2}), or by the subsequent scattering that brings the system to
thermal equilibrium.  In the Fermi-degenerate regime, we know that $e(N,T)
\approx (3/5) \epsilon_{F} (N) + (\pi^{2}/4) [(k_{B}^{2}
T^{2})/\epsilon_{F}(N)]$,
correct up to order $[T/T_{F}]^{2}$.  Equating $e(N,T)$ to its initial
value $e(N_{0},T_{0})$, we solve for $T$.  
Denoting the initial Fermi-temperature by $T_{F,0}$,
using that
$[\epsilon_{F}(N_{0})/\epsilon_{F}(N)] = [N_{0}/N]^{2/3}$, and
$[T_{F}/T_{F,0}] = [N/N_{0}]^{2/3}$, we obtain
\begin{eqnarray}
\frac{T}{T_{F}} =&& \left(
\frac{T_{0}}{T_{F,0}} \right) \left( \frac{N}{N_{0}} \right)^{-1/3} 
\; \times
\nonumber \\
&& \sqrt{ 1 + \frac{12}{5 \pi^{2}} \left[ \frac{T_{F,0}}{T_{0}}
\right]^{2} \left[ 1 - \left( \frac{N}{N_{0}} \right)^{2/3} \right] } \; .
\label{e:hh8}
\end{eqnarray}
To obtain a time-dependent expression, we substitute 
the fraction of remaining particles
by $N(t)/N_{0}  = \exp(-t/\tau_{L}), [1 + t/\tau_{L}]^{-1},
[1+ 2 t/\tau_{L}]^{-1/2}$ for loss that is predominantly
caused by background, two- and three-body scattering respectively.
In Fig.(1), we graph the time-dependent temperature to
Fermi-temperature ratio for $T_{0} = 0.01 T_{F}$.
The short-time behavior, $t < 0.1 \tau_{L}$, is indeed well
described by Eq. (\ref{e:hh7}).  On the scale of the graph ($t \sim 
\tau_{L}$, $T \sim 0.1 T_{F}$) the plot is `universal' in the sense
that any temperature curve of lower initial value is 
indistinguishable from that shown (described by the $T_{0}
\rightarrow 0$--limit, $T \rightarrow T_{F,0}
\sqrt{ (12/5\pi^{2}) [(N/N_{0})^{-2/3} -1 ] }$).  After $t \approx
0.1 \tau_{L}$, the temperature increase slows down and $T$ reaches
$25 \% - 30\%$ of the Fermi-temperature by `middle age',
$t \sim 0.5 \tau_{L}$.  At later times, the fermion system becomes
classical (invalidating the above expression of $e(N,T)$) and
$\lim_{t \rightarrow \infty} T(t) = 0.4 T_{F,0}$,
which follows from $e_{classical} = (3/2) k_{B} T \approx 
(3/5) \epsilon_{F}(N_{0})$.  The
above result does not depend on thermal equilibrium, provided
we interpret temperature as a measure of the difference of 
the system's energy with its ground state value.

\underline{Superfluid formation and heating as competing effects}
Regardless of the initial temperature,
the fermions heat up to at least $10 \%$ of the Fermi-temperature
after only $7 \%$ of the system's lifetime.  The temperature
$T = 0.1 T_{F}$
is the highest possible critical temperature for fermion superfluidity,
$T_{c,max} \approx 0.1 T_{F}$, therefore,
the maximal time the fermion superfluid has to form
is $0.07 \tau_{L}$, not $\tau_{L}$. 
It may, in fact, be difficult to realize $T_{c,max}$ 
and we consider the more general constraints on the
interaction from the requirement of forming
the Cooper-paired superfluid before the system heats up above the
critical temperature:
\begin{equation}
T(t=\tau_{form}) < T_{c} \; .
\label{e:hh9}
\end{equation}
Because the superfluid formation time cannot exceed
$0.07 \tau_{L}$, $T_{0} \sqrt{1 + 3 t/\tau_{2}}$ (\ref{e:hh7}) adequately
describes the temperature, and we find from (\ref{e:hh9})
\begin{equation}
\tau_{2} > \frac{3 \tau_{form}}{(T_{c}/T_{0})^{2} -1} \; .
\label{e:hh10}
\end{equation}
Substituting $\tau_{2}$ and ${\tau_{form}}$ in terms of $\tau_{L}$
(\ref{e:hh7}) and $\tau_{Fermi}$ (\ref{e:hh1b}), 
the inequality (\ref{e:hh10}) yields
\begin{equation}
\tau_{L} > \left( \frac{4}{5 \pi^{3}} \right) \;
\frac{ (T_{F}/T_{c})^{4} }{ 1 - (T_{0}/T_{c})^{2}} \;
\tau_{Fermi} \; .
\label{e:hh11}
\end{equation}
If the experimentalist succeeds in bringing
the initial temperature of the system significantly below
the critical temperature (e.g. $T_{0} \leq 0.3 T_{c}$),
then the denominator in (\ref{e:hh11}) can be replaced 
by $1$ and the condition becomes independent of $T_{0}$.  
Assuming $T_{0} \leq 0.3 T_{c}$ and substituting $T_{c}/T_{F}$
by $0.3 \exp(-0.4 r/|a|)$, we finally obtain
\begin{eqnarray}
\frac{|a|}{r} &>& \frac{0.4}{
\ln[0.3/(4/5\pi^{3})^{1/4}] + 
(1/4) \ln(\tau_{L}/\tau_{Fermi})  } 
\nonumber \\
&& \approx 
\frac{0.69}{log_{10}(\tau_{L}/\tau_{Fermi})} \; ,
\label{e:hh12}
\end{eqnarray}
where the last approximation is valid if $\tau_{L} \geq
10^{4} \tau_{Fermi}$.  Note that the minimal value of the
dimensionless interaction strength $|a|/r$ is approximately
twice as large as from the condition $\tau_{form} > \tau_{L}$.
Thus, if $\tau_{L}/\tau_{Fermi} \sim 10^{6} - 10^{7}$, the
minimal value the negative scattering length has to take on
to form Cooper-pairs is, $|a|/r > 0.125 - 0.106$.  

	Equation (\ref{e:hh9}) tacitly assumes that Fermi-holes
relax faster than the superfluid can form, an assumption we
now justify by calculating the lifetime of a hole.  Specifically,
we consider a fermion 1 hole in the center of the Fermi-sphere
$(1,{\bf k}=0)$, and we determine its inverse lifetime
$\tau_{h,{\bf k}=0}^{-1}$ in a Fermi-Golden rule calculation as
the rate at which the hole `fills up'' by scattering processes
$(1,{\bf q}) + (2,{\bf p}) \rightarrow (1,{\bf k}=0) + (2,{\bf p}
+ {\bf q})$:
\begin{equation}
\tau_{h,{\bf k}=0}^{-1} = \left( \frac{2 \pi}{\hbar} \right)
\left[ \frac{4 \pi \hbar^{2}}{m a \Omega} \right]^{2}
\sum_{{\bf p},{\bf q}} n_{1,{\bf q}} n_{2,{\bf p}}
(1 - n_{2,{\bf p}+{\bf q}}) \delta (E_{initial} - E_{final})
\label{e:extra}
\end{equation}
where $\Omega$ denotes the volume in box normalization.
With the $(1,{\bf k}=0)$ final state, the energy density
reduces to $\delta (E_{initial} - E_{final}) \rightarrow
(m/\hbar^{2} p q) \delta (\cos\theta)$, where $\theta$ denotes
the angle between ${\bf q}$ and ${\bf p}$.  The (${\bf q}$,
${\bf p}$)--integral with zero-temperature Fermi Dirac
occupation numbers yields $\tau_{h,{\bf k}=0}^{-1}
= (4 \pi a^{2}) n \left( \frac{3 \hbar k_{F}}{4 m} \right)$.
Remarkably, $\tau_{h,{\bf k}=0}^{-1}$ equals the classical
estimate for the collision rate of a zero velocity particle
in the presence of the distinguishable particles, $\tau_{coll}^{-1}$
where $\tau_{coll} = [\langle \sigma n v \rangle]^{-1} 
= 0.207 \tau_{Fermi} (r/|a|)^{2}$.  Replacing the Pauli-blocking
factor $(1 - n_{2,{\bf p}+{\bf q}})$ in (\ref{e:extra}) by
1 reduces the lifetime to half its value.  Therefore, whereas
Pauli-blocking increases the lifetime of holes and particles
drastically near the Fermi-surface, its effect on the hole lifetime
is moderate near the center of the Fermi-spehere where
$\tau_{h,{\bf k}} \sim \tau_{coll}$, which is also the time scale
on which the high energy $(2,{\bf p}+{\bf q})$--fermions scatter.
Finally, $\tau_{form}/\tau_{coll} \approx 7.7 (|a|/r)^{2} 
\exp(0.8 \frac{r}{|a|})$, so that at its minimum 
($|a| = 0.4 r$) $\tau_{form}/\tau_{coll} \approx 9.5$
and $\tau_{form}/\tau_{coll}$ quickly
increases as $|a|$ is reduced.  The implication is that
that there is a region inside the
Fermi-sphere where holes are short-lived on the
time scale of superfluid formation.  The decay of those holes
produces `high energy' fermions that start heating the
system while the superfluid forms.

	While the treatment (\ref{e:hh9})--(\ref{e:hh12}) 
certainly simplifies much of the
intricate dynamics, we can trust the accuracy of
the final constraint (\ref{e:hh12}) to at least
$15 \%$.  Consider, 
for instance, the possibility that the non-equilibrium
nature of the formation process, or the change of 
$c_{V}$ as the superfluid is formed, slows down the
heating or favors the superfluid creation so as
to weaken the condition (\ref{e:hh9}) effectively
by as much as one order of magnitude $T(t=\tau_{form}/10) < T_{c}$.
This amounts to multiplying $\tau_{L}$ by a factor
of 10, giving, under the same conditions
as above a range of minimal values $|a|/r \sim 0.106 - 0.092$.

	Finally, we claim that our estimates are lower
bounds.  In a spherically symmetric harmonic trap
the same reasoning as for the homogeneous mixture leads 
to the following time-dependence for the temperature
as a result of background scattering:
$T/T_{F} = [T_{0}/T_{F,0}] (N/N_{0})^{-1/6}
\sqrt{ 1 + (3/\pi^{2}) (T_{F,0}/T_{0})^{2} [ 1 - (N/N_{0})^{1/3}] }$
\cite{harm},
with $N/N_{0} = \exp(-t/\tau_{L})$. On the other hand,
two- and three-body rates are highest in the high-density middle of the 
trap where the energy per particle is lowest.  
Thus, lower energy atoms are removed
preferentially \cite{hulet3}, giving a temperature
that increases more rapidly than in the homogeneous system.  
Secondly, rates for some recombination processes 
are sensitive to the value of the scattering length.  A large 
value of $|a|$ can imply a shorter
lifetime, in which case $\tau_{L}$ in (\ref{e:hh12}) 
really depends on $|a|$ \cite{Fed}, \cite{cg}.

        In conclusion, we have determined how a metastable
degenerate fermion gas heats from the hole creation
caused by the loss processes.  The heating rate is particularly
significant at ultra-cold temperatures, with a 
temperature doubling time, $\tau_{2} \sim 20
[T_{0}/T_{F}]^{2} \tau_{L}$.
On longer time scales $t \sim 0.5 \tau_{L}$, the fermion
system heats up to $T \geq 0.25 T_{F}$.  For the quest of atom-trap
fermion superfluidity, this heating competes with
the formation of the superfluid and we derive a minimal
value for the unlike-fermion scattering length to
observe superfluid formation in a gas mixture of unlike
fermion atoms.  In contrast to the simple mixture of
fermion atoms, sympathetic cooling can provide
a third (bosonic) atom species that continues to absorb the energy released by
hole creation.  Given the importance of the hole-heating
mechanism, such continuous cooling may be
helpful to reach fermion superfluidity.

        In a question at a Los Alamos colloquium, 
Albert Pecheck remarked upon the problem of Fermi-holes.  The fact 
that this problem did not appear to have been addressed,  
motivated this work.  The author also gratefully acknowledges 
discussions on this subject with Peter Milonni, Kyoko Furuya,
Randy Hulet, Xinxin Zhao and Michael Di Rosa.

          
\newpage
\noindent \underline{Fig. 1} Plot illustrating the loss-induced heating
	  of the fermion mixture.  For a homogeneous fermion gas mixture
	  with initial temperature of
	  $T = 0.01 T_{F}$, we plot the ratio of temperature
	  to fermi-temperature as a function of time for loss processes
	  that are predominantly background
	  (solid line), two-body (dashed line) and three-body (long-dashed 
	  line) processes \cite{remark}. The thin line curve shows 
	  $T/T_{F}$ for the short time temperature approximation, 
	  $T = T_{0} \sqrt{1+3t/\tau_{2}}$
	  in the case of background scattering.



\begin{thebibliography}{99}
\bibitem{remark} Barring unusual circumstances, the
lifetime $\tau_{L}$ due to inelastic scattering
of the particle is effectively  energy independent.
\bibitem{harm} A lengthy but straightforward calculation gives
the following expression for the energy per particle in
a spherically symmetric harmonic trap of degenerate fermion
particles:$ e = (3/4) \epsilon_{F} [ 1 + (T/T_{F})^{2} \pi^{2}/3]$
where $\epsilon_{F}$ is the Fermi-energy, $\epsilon_{F} = 
\hbar \omega_{T} (6N)^{1/3}$, and where $N$ denotes the total
number of fermions and $\omega_{T}$ the trap frequency.
\end{thebibliography}
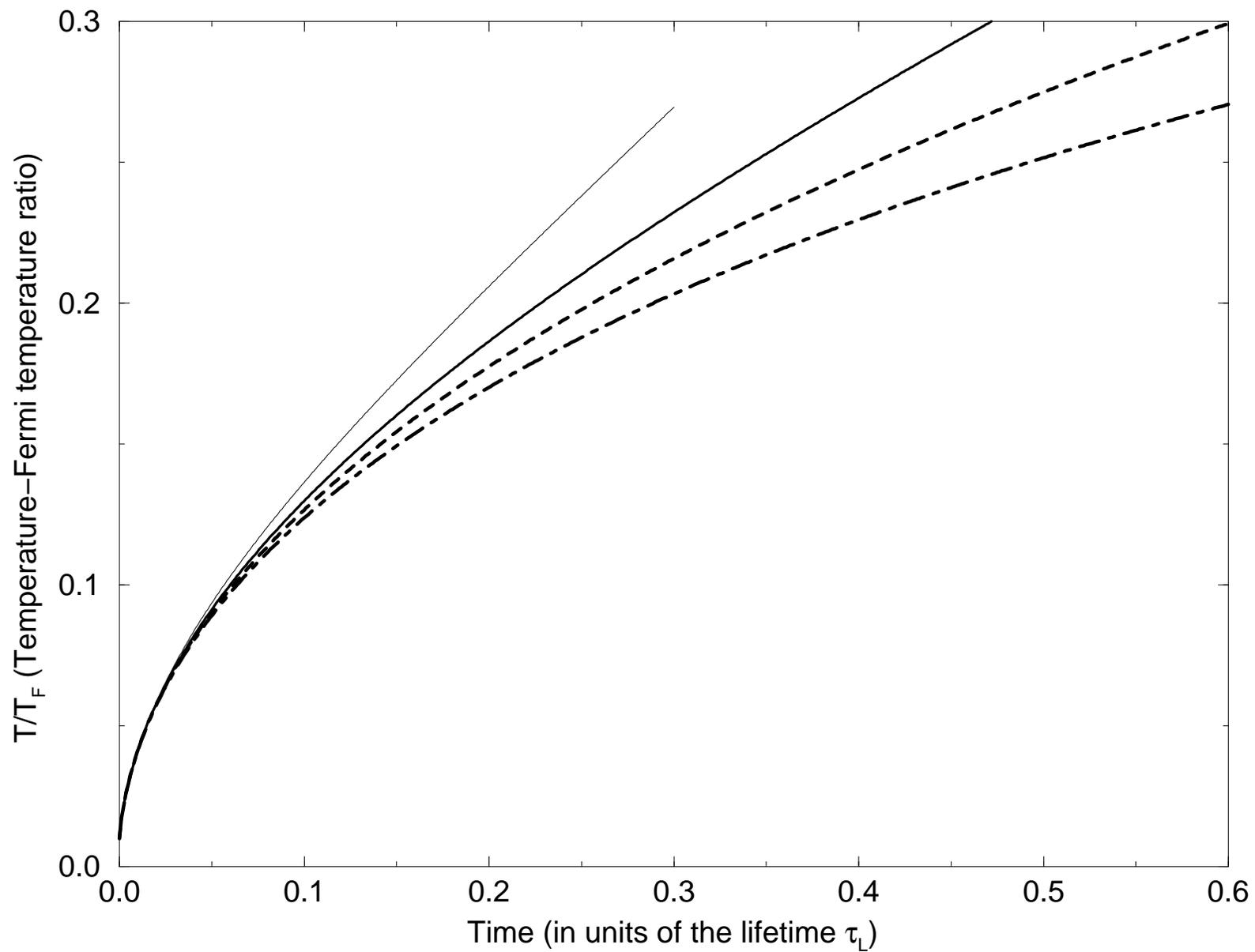
\end{document}